\documentclass[aps,prl,twocolumn,superscriptaddress,floatfix,longbibliography]{revtex4-1}
\usepackage{graphicx}
\usepackage{color}
\usepackage{float}
\usepackage{rotating}

\newcommand{\dR}{$\Delta$$R$/$R$}
\newcommand{\Odd}{O$_{d\text{-}d}$}
\newcommand{\Tdd}{T$_{d\text{-}d}$}

\newcommand{\FMO}{Fe$_{2}$Mo$_{3}$O$_{8}$}

\newcommand{\FMOX}{(Fe$_{1-x}$Zn$_{x}$)$_{2}$Mo$_{3}$O$_{8}$}

\begin{document}

\title{Picosecond creation of switchable optomagnets from a polar antiferromagnet with giant photoinduced Kerr rotations}

\author{Y. M. Sheu}
\email{ymsheu@nctu.edu.tw}
\affiliation{Department of Electrophysics, National Chiao Tung University, Hsinchu, 30010, Taiwan}
\affiliation{Center for Emergent Functional Matter Science, National Chiao Tung University, Hsinchu 30010, Taiwan}
\author{Y. M. Chang}
\affiliation{Department of Electrophysics, National Chiao Tung University, Hsinchu, 30010, Taiwan}
\author{C. P. Chang}
\affiliation{Department of Electrophysics, National Chiao Tung University, Hsinchu, 30010, Taiwan}
\author{Y. H. Li}
\affiliation{Department of Electrophysics, National Chiao Tung University, Hsinchu, 30010, Taiwan}
\author{K. R. Babu}
\affiliation{Department of Physics and Center for Theoretical Physics, National Taiwan University, Taipei, 10617, Taiwan}
\affiliation{Physics Division, National Center for Theoretical Sciences, Hsinchu 30013, Taiwan}
\author{G.Y. Guo}
\affiliation{Department of Physics and Center for Theoretical Physics, National Taiwan University, Taipei, 10617, Taiwan}
\affiliation{Physics Division, National Center for Theoretical Sciences, Hsinchu 30013, Taiwan}
\author{T. Kurumaji}
\affiliation{Department of Physics, Massachusetts Institute of Technology, Cambridge, MA 02139-4307, USA}
\author{Y. Tokura}
\affiliation{RIKEN Center for Emergent Matter Science (CEMS), Wako, Saitama 351-0198, Japan}
\affiliation{Department of Applied Physics and Tokyo College, University of Tokyo, Tokyo 113-8656, Japan }
\begin{abstract}
On-demand spin orientation with long polarized lifetime and easily detectable signal is an ultimate goal for spintronics. However, there still exists a trade-off between controllability and stability of spin polarization, awaiting a significant breakthrough. Here, we demonstrate switchable optomagnet effects in \FMOX, from which we can obtain tunable magnetization, spanning from -40$\%$ to 40$\%$ of a saturated magnetization that is created from zero magnetization in the antiferromagnetic state without magnetic fields. It is accomplishable via utilizing circularly-polarized laser pulses to excite spin-flip transitions in polar antiferromagnets that have no spin canting, traditionally hard to control without very strong magnetic fields. The spin controllability in \FMOX~ originates from its polar structure that breaks the crystal inversion symmetry, allowing distinct on-site $d$-$d$ transitions for selective spin flip.  By chemical doping, we exploit the phase competition between antiferromagnetic and ferrimagnetic states to enhance and stabilize the optomagnet effects, which result in long-lived photoinduced Kerr rotations. The present study, creating switchable giant optomagnet effects in polar antiferromagnets, sketches a new blueprint for the function of antiferromagnetic spintronics.
\end{abstract}

\maketitle

Historically, antiferromagnetic materials have long been disregarded since the lack of a macroscopic magnetization renders their controllability difficult. However, there is a vast variety of spin arrangements in antiferromagnets formed thermodynamically to resist electronic or magnetic perturbations, ideal for maintaining spin orientations. As they are so robust against external perturbations to alter spin configurations, it is challenging to obtain useful properties from antiferromagnets.  Some successful cases, including antiferromagnetic (AFM) heterostructures \cite{Meiklejohn1957PR,Parkin1991APL,Schulthess1998PRL,Nogues1999JMMM}, AFM multiferroics \cite{Kimura2003Nature,Katsura2005PRL,Ramesh2007NM,Chu2008NMat,Sheu2014PRX}, techniques like optical switches \cite{Kimel2005Nature,Li2013Nature,Nemec2018NP}, and so on \cite{Jungwirth2016NNano}, have realized potential AFM applications by taking a detour to search for tunable or switchable magnetoresistance, spin induced ferroelectricity, or AFM resonance mode generation. Moreover, while methods exploiting spin transfer torques have been frequently applied to directly flip spin in AFM metal films, they show small effects and suffer from stability problems \cite{MacDonald2011}. Despite the enormous effort made by researchers, the trade-off between controllability and stability still exists and the sign of spin polarization (polarity) remains uncontrollable.

\begin{figure}[h!]
\begin{center}
\includegraphics[width=3in]{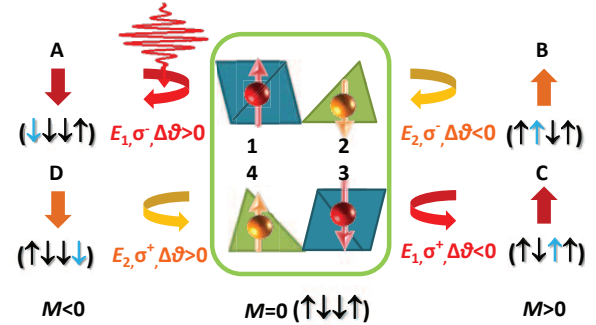}
\caption{ \label{fig0} Illustration of optomagnets created by optical helicities in a magnetic basis. Four distinct spin quantum states, A, B, C, D, can be selectively created from the zero-magnetization ($M$=0) state by flipping one of the sublattice spin moments (1,2,3,4)=($\uparrow\downarrow\downarrow\uparrow$), with the flipped moment color coded in each unique configuration.  This can be made possible through the on-site $d$-$d$ optical transitions with combinations of pump photon energies, $E_{1,2}$, and circularly-polarized pumps, $\sigma^{\pm}$. Combining the resultant magneto-optical Kerr rotations, $\Delta\theta$, we can unambiguously discern the four distinct microscopic states that lead to two switchable magnetization direction macroscopically.    }
\end{center}
\end{figure}

In magnetic oxides, the degeneracy of the five $d$ orbitals is lifted due to the imposed crystal field from oxygen cages, potentially allowing spin-flip transitions. However, an on-site $d$-$d$ optical transition is both spin and parity forbidden in the local centrosymmetric environment. Fortunately, this spin-flip transition can be allowed and enhanced due to crystal distortions, ionic interactions, spin-orbit interactions and so on. As the oscillator strength strongly depends on the participating states, the most efficient way to enhance the on-site $d$-$d$ transition is to have an odd parity crystal field term, e.g., $ax+by+cz$, arising from a crystal distortion \cite{Kahn1969PR}. Therefore, having a polar structure can significantly enhance the on-site $d$-$d$ transitions. Since most of AFM materials remain centrosymmetric, rare studies have employed this method to control spin polarity. Despite that the on-site $d$-$d$ transition may be allowed in AFM materials, however, strong spin exchange interactions may destabilize the excited states, likely leading to fast decay of the photo-generated spin-polarized states.

Here, we explore polar antiferomagnets without an inversion symmetry that allows spin flip transitions on the magnetic ions. We exploit ultrashort laser pulses to flip sublattice spin in AFM ordered multiferroic \FMO, creating four distinct on-demand spin configurations in a magnetic basis illustrated in Fig. \ref{fig0}, which can be selected by combinations of photon energies, $E_{1,2}$, and optical helicities, $\sigma^{\pm}$, enabling to produce a switchable magnetization measured via magneto-optical Kerr rotations $\Delta\theta$. In this article, the (optical) helicities specifically refer to the case of circularly-polarized light, $\sigma^{\pm}$. We further demonstrate that if the ground state antiferromagnetism is competing with a ferrimagnetic (FRM) ordering, such optomagnet effects can be energetically stabilized with the enhancement of signals by up to two orders of magnitude accompanying with much a prolonged magnetization lifetime, which is accomplishable through chemical doping in \FMO~\cite{Kurumaji2015PRX}. Our demonstration utilizing antiferromagnets, which overcome the trade-off between controllability and stability, may open new routes for operating spintronics on demand, e.g., optomagnet functions that turn on a magnetization with switchable polarities set by optical helicities, Fig. \ref{fig0}, for operating on/off exchange couplings to other spin-based devices.

\begin{figure}[tb]
\begin{center}
\includegraphics[width=3.45in]{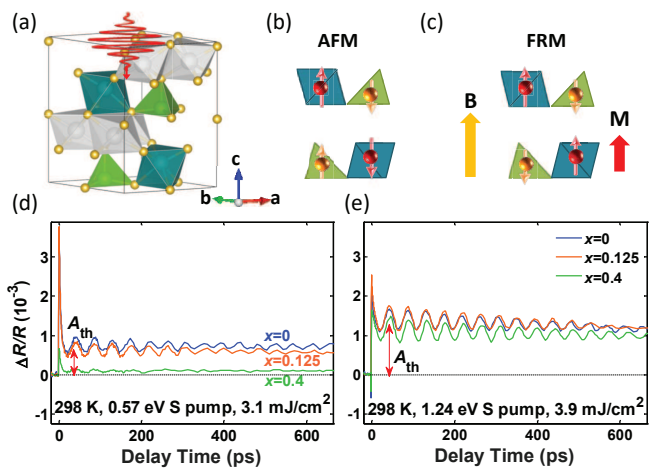}
\caption{ \label{fig1}(a) A Schematic diagram for the unit cell of \FMO~ with Fe$^{2+}$ ions in the green shaded oxygen octahedrons (O-site) and blue shaded oxygen tetrahedrons (T-site). The Mo$^{4+}$ ions are located within the grey shaded oxygen octahedrons. (b,c) Each illustration displays four sublattices in a magnetic basis for the antiferromagnetic (AFM) phase, (b), and ferrimagnetic (FRM) phase, (c).  (d,e) Time-resolved reflectivity change, \dR, measured at room temperature with probe light centered at $\sim$0.86 eV. The absorption, $A_{\text{th}}$ (see text for the definition),  is comparable for different Zn concentrations, $x$ in \FMOX, when the pump photon energy is above $\sim$1 eV. However, $A_{\text{th}}$ is greatly reduced with excitation energies $\sim$0.5-0.6 eV for a $x$=0.4 sample, which replaces 80$\%$ of Fe$^{2+}$ ions on the  T-site. The oscillatory behavior in \dR~is due to the Brillouin scattering of the probe light, arising from a propagation of coherent acoustic strain generated by laser heating \cite{Thomsen1986PRB}.
}
\end{center}
\end{figure}

\FMO~ is pyroelectric (space group $P6_{3}mc$) at room temperature and magnetically ordered below transition temperature $\sim$60 K \cite{Bertrand1975JPF,LePage1982ACSB}. The oxygen cage of Fe$_{2}$ (magnetic) layer and Mo$_{3}$ (nonmagnetic) layer are stacking alternatively along the $c$-axis, Fig. \ref{fig1}(a). The spin moment of intralayer Fe$^{2+}$ on the oxygen octahedron (O-site) and on the oxygen tetrahedron (T-site) are in AFM arrangement.  In the AFM ground state, the interlayer magnetic arrangement between the same sites (O-O and T-T) is AFM, while that between different sites (O-T) is ferromagnetic (FM), Fig. \ref{fig1}(b).  The application of a magnetic field ($\|c$) turns the AFM ground state to the FRM state, and the nearest interlayer magnetic configuration for O-T is AFM while that for O-O and T-T is FM (Fig. \ref{fig1}(c)). This magnetic transition is of the first order arising from the AFM and FRM phase competition, which is also controllable by energetic Zn substitutions on T-sites, \FMOX, that lower the free energy difference between the two phases by reducing the AFM interaction between T-sites \cite{Bertrand1975JPF,Kurumaji2015PRX}.

\begin{figure*}[tb]
\begin{center}
\includegraphics[width=7in]{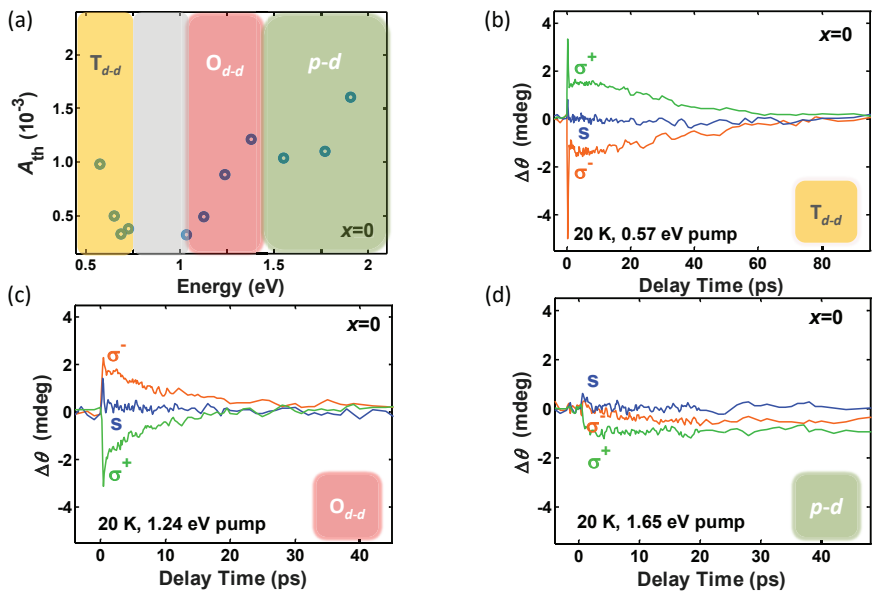}
\caption{\label{fig2} (a) Absorption spectrum $A_{\text{th}}$ as a function of photon energy at room temperature. (b,c,d) Time-resolved magneto-optical Kerr effect (TRMOKE) measured at different excitation energies with various polarizations. TRMOKE displays two distinct characters for (b), $d$-$d$ transition on T-site (\Tdd) around 0.5-0.6 eV and (c), $d$-$d$ transition on O-site (\Odd) near 1.1-1.5 eV.  Both excitations show circularly-polarized pump ($\sigma^{\pm}$) dependence, which disappears for $p$-$d$ charge transfer excitation ($p$-$d$) above 1.55 eV, with excitation at 1.65 eV displayed in (d). All the data shown here were measured from the x=0 sample.
}
\end{center}
\end{figure*}

Due to the inherent crystal polar structure of \FMOX, the inversion symmetry is broken, allowing the originally dipole-forbidden $d$-$d$ transitions. We use absorption induced laser heating effect, $A_{\text{th}}$ indicated in the time-resolved reflectivity change \dR~ of Figs. \ref{fig1}(d) and \ref{fig1}(e), to display the absorption spectrum, Fig. \ref{fig2}(a). The $d$-$d$ transition of Fe$^{2+}$ on T-sites (\Tdd) is around 0.5-0.6 eV, and the absorption reduces with the increment of Zn replacing Fe on T-site, Fig. \ref{fig1}(d). The absorption diminishes between 0.6-1 eV (Fig. \ref{fig2}(a)), above which \FMOX~becomes very opaque and displays large \dR~signals, Fig. \ref{fig1}(e) and Fig. \ref{fig2}(a). The $d$-$d$ transition of Fe$^{2+}$ on O-sites (\Odd) and the $p$-$d$ charge transfer transition are experimentally discerned via time-resolved magneto-optical Kerr effects (TRMOKE) measured from \FMO~ in the AFM ground state, Figs. \ref{fig2}(b)-\ref{fig2}(d). The helicity-dependent ($\sigma^{\pm}$) TRMOKE signal disappears above 1.55 eV, i.e., the $p$-$d$ charge transfer region, below which the onset of \Odd~is located. Both \Odd~and \Tdd~ display significant TRMOKE signals when the circularly-polarized pump is employed. The two distinct $d$-$d$ transition energies agree with the crystal field splitting of Fe$^{2+}$ on O-site and on T-site \cite{Varret1972JPF,Slack1966PR}, and our ab initio calculation \cite{suppl} of the absorption also shows quantitative consistency with the experimental assignments. It should be noted that while only the absorption spectrum of the $x$=0 sample is displayed in Fig. \ref{fig2}(a), the $A_{\text{th}}$ signals from the samples of $x$=0.125 and $x$=0.4 are nearly the same as that from the $x$=0 sample between 1-2 eV at room temperature. Below 1 eV, only the $x$=0 and $x$=0.125 samples display similarity in transient reflectivity, whereas the $x$=0.4 sample barely shows measurable signals, Fig. \ref{fig1}(d).

\begin{figure*}[t]
\begin{center}
\includegraphics[width=7in]{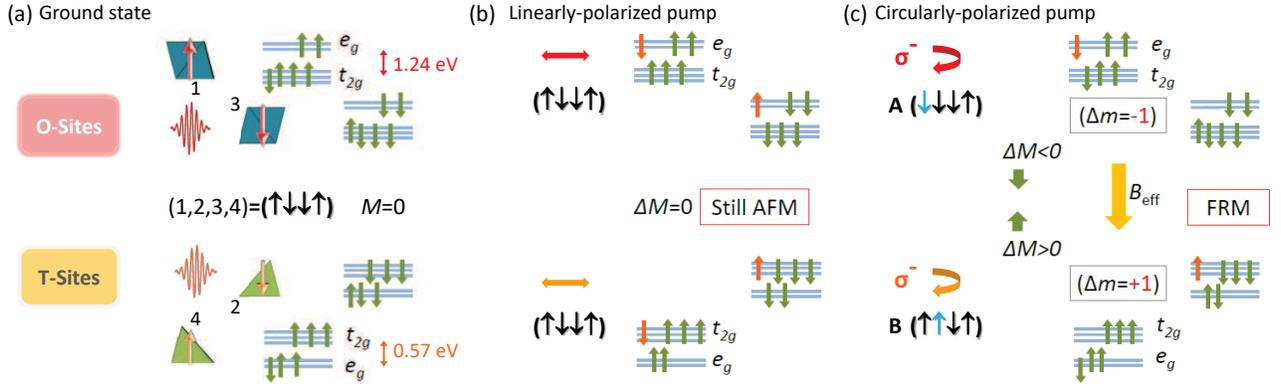}
\caption{\label{fig3} Magnetization induced from the selective sublattice excitation.  The upper (lower) schematics are displayed for O-sites (T-sites). (a) Ground state energy levels for O-sites (T-sites) on sublattice 1 and 3 (2 and 4). The $d^{6}$ electrons are displayed with the splitted energies indicated by the double arrows. The energy separation is derived from measurements of Fig. \ref{fig2}, in consistent with the 10 Dq of cubic field \cite{Varret1972JPF,Slack1966PR}. (b) The excited state created from linearly-polarized light. The excited spin moments (orange arrows) preserve the total magnetic moments on both O- and T-sites, i.e., $\Delta M$=0. (c) The excited state created from circularly-polarized photons, $\sigma^{-}$ shown here. The transition involves the spin selection rule, i.e., $\Delta$$m$=$\pm$1, with the flipped moment (orange arrows) frustrating the zero magnetization, i.e., $\Delta M\neq$0. See text for further detailed discussion.  Here, the two transitions demonstrate the mechanism for a magnetic basis in Fig. \ref{fig0} to create the configurations A and B from the sublattice spin moments ($\uparrow\downarrow\downarrow\uparrow$) of $M$=0.
}
\end{center}
\end{figure*}

A magneto-optical polar Kerr rotation originates from a circular birefringence of materials with a magnetization $M$. The light propagating along $M$ will rotate its polarization upon the reflection. When the $M$ direction inverts, the Kerr rotation also reverses, irrespective of absorption strength in the optical and near IR region \cite{Sato1983JSAP}. The polarization reversal of light can only be induced from the magnetization inversion due to the broken time-reversal symmetry. None of these characters were ever observed in collinear antiferromagnets due to the zero magnetization \cite{Feng2015PRB}.  Therefore, the reversible TRMOKE signal created from the circularly-polarized pump after excitation of \Tdd~(Figs. \ref{fig2}(b)) or \Odd~(Figs. \ref{fig2}(c)) in the AFM ordered \FMO~ displays the switchable photoinduced magnetization, i.e., optomagnets, which have to involve a spin-flip process during the $d$-$d$ transition. More insight can be gained from the observation of the opposite Kerr rotations, which arise from \Odd~ and \Tdd~ while the same pump helicity is employed, the orange-line pair (or the green-line pair) in Figs. \ref{fig2}(b) and \ref{fig2}(c), pointing to AFM coupling as the strongest interaction between Fe$^{2+}$ on O-sites and Fe$^{2+}$ on T-sites. As a result, opposite magnetic moment switches with the same pump helicity applied from \Odd~and \Tdd., analogous to the magnetic-field induced reversal. What differs from the field switch is that only the pump photon-energy selected sublattice can flip spin moments, while spin on both sites responds to the magnetic field ($\|c$) simultaneously, Fig. \ref{fig1}(c).

From the absorption spectrum and TRMOKE signals, we have already attributed the two absorption peaks below 2 eV to the \Odd~ and \Tdd. Figure \ref{fig3}(a) displays the distinct sites for these excitations with the associated energy levels of the ground state. The two transitions arise from the splitted $t_{2g}$ and $e_{g}$ levels caused by the different crystal fields with the energy separation of $\sim$1.24 eV for O-site and  $\sim$0.57 eV for T-site, upper and lower panel of Fig. \ref{fig3}(b) \cite{CFNote}, respectively. Therefore, the spin selection rule has to be employed for both transitions, i.e., the spin moment remains the same after an absorption of linearly-polarized light (Fig. \ref{fig3}(b)) whereas it flips after an absorption of $\sigma^{\pm}$ (Fig. \ref{fig3}(c)). Using up ($\uparrow$) and down ($\downarrow$) spin moments to represent the individual sublattice moment in the magnetic basis illustrated in Fig. \ref{fig0}, we can obtain a configuration ($\uparrow\downarrow\downarrow\uparrow$) for the AFM ground state with the spin moments located on sublattices (1,2,3,4) in Fig. \ref{fig0}. Because the TRMOKE signal shows rotation reversal under the same helicity pump from different excitations, e.g., $\sigma^{-}$ from \Odd~and \Tdd, we illustrate this discrepancy in Fig. \ref{fig3}(c). That is to say, the same circularly-polarized pump of \Odd~ and \Tdd~ will excite opposite spin moments of O-site and T-site, e.g., sublattices 1 and 2 will be excited by $\sigma^{-}$ pump of $\sim$1.24 eV and $\sim$0.57 eV respectively, thus frustrating the compensated spin moments to induce a transient magnetization. Such an opposite response of spin moment to the same circularly-polarized light on O-site and T-site can be viewed as the effective magnetic field, $B_{\text{eff}}$, acting on the specific sublattice as illustrated in Fig. \ref{fig3}(c). Discrimination between $B_{\text{eff}}$ and general magnetic field originates from the sublattice selectivity.

Employing the above mechanism, we can demonstrate a scheme to create optomagnets in the AFM phase by using a circularly-polarized pump, as displayed in Fig. \ref{fig0}.  Four distinct sublattice spin quantum states can be initiated by different pump excitations and regarded as A($\downarrow\downarrow\downarrow\uparrow$), B($\uparrow\uparrow\downarrow\uparrow$), C($\uparrow\downarrow\uparrow\uparrow$), and D ($\uparrow\downarrow\downarrow\downarrow$), selectively created from (\Odd, $\sigma^{-}$) (upper panel in Fig. \ref{fig3}(c)), (\Tdd, $\sigma^{-}$) (lower panel in Fig. \ref{fig3}(c)), (\Odd, $\sigma^{+}$), and (\Tdd, $\sigma^{+}$) respectively, leading to the transient Kerr rotations observed in Figs. \ref{fig2}(b) and \ref{fig2}(c) and the associated optomagnets illustrated in Fig. \ref{fig0}. We can see that writing with four combinations of helicity/energy and reading from Kerr rotations may have potential applications for an AFM based data storage via the optomagnet effect.

We note that the photoinduced magnetization and the associated optomagnet here differs from other photoinduced AFM-FM phase transitions \cite{Ju2004PRL,Thiele2004APL} in two major ways. First, the photoinduced AFM-FM phase transition as reported so far is initiated by laser heating to cross the phase transition temperature, while the optomagnet described here arises from a direct spin-flip transition right after the absorption. Second, the photoinduced AFM-FM phase transition lacks the polarity choices; however, the optomagnet effect can select a polarity on demand. Thus, the optomagnets turned on by a circularly-polarized pump is distinct from other heating induced phase transitions.

\begin{figure}[tb]
\begin{center}
\includegraphics[width=3.45in]{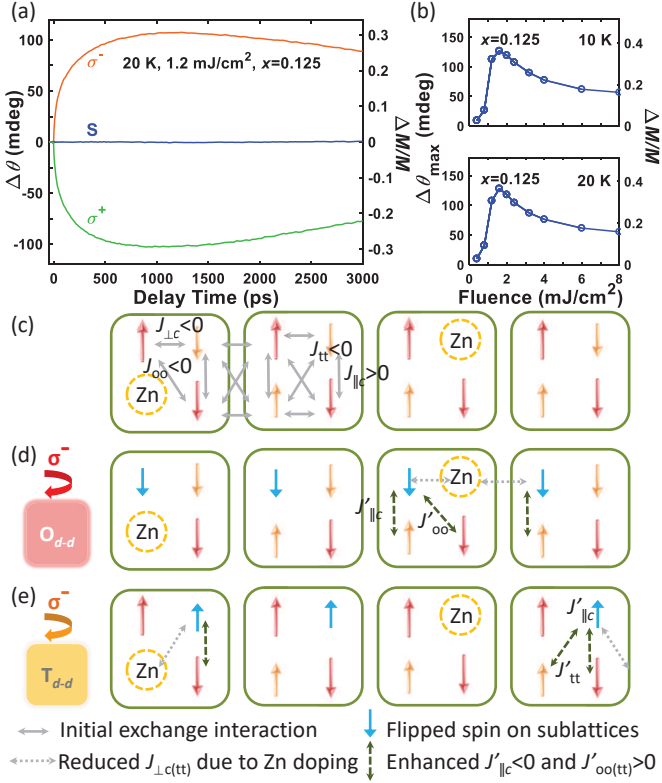}
\caption{\label{fig4} (a) Large Kerr rotation developed for the $x$=0.125 with a circularly-polarized pump from \Odd. (b) Nearly identical fluence dependence for the maximum TRMOKE signal at 10 K and 20 K. In both (a) and (b), the left axis displays the photoinduced Kerr rotation angle, whereas the right axis shows the ratio of photoinduced magnetization to the saturated magnetization. (c) Four magnetic bases are displayed to show the homogeneous Zn substitution for the $x$=0.125 sample. The exchange couplings are depicted with grey double arrows in the first two bases. (d,e) Excited states for \Odd~ and \Tdd. The removal of AFM coupling between the intralayer (interlayer) O-T (T-T) interaction for \Odd~(\Tdd) is indicated by the dotted grey double arrows. The altered exchange interaction, $J'_{ij}$, after excitations is displayed in dashed green double arrows. The stabilization after circularly-polarized photoexcitation, $\sigma^{-}$ here, can be achieved through these modulations of exchange interactions (see text for the detailed discussion).  }
\end{center}
\end{figure}

Further investigating the optomagnet effect on Zn doped samples, we observe giant Kerr rotations for the $x$=0.125, Fig. \ref{fig4}(a). The maximum Kerr rotation ($\Delta\theta_{\text{max}}$) induced by a circularly-polarized pump is nearly two-order-of-magnitude larger than that for the $x$=0 sample. The large magneto-optical response can be induced by relatively low pump flunece, Fig. \ref{fig4}(b), that does not significantly increase average lattice temperature, e.g., an average temperature elevation would be $\sim$2 K for excitation fluence around 1 mJ/cm$^2$, which can be further reduced with a design of thin film heterostructure \cite{Sheu2018APL}. Moreover, the optomagnet effect for the $x$=0.125 sample can persist for more than several nanoseconds to less than a quarter millisecond, i.e., relaxation occurs before the next laser pulse excitation, $\sim$200 $\mu$s, with switchable direction selectable by a circularly-polarized pump. The giant photoinduced Kerr rotation disappears when the crystal undergoes the transition to the paramagnetic state above 48 K, see Fig. S6 in the supplemental material. Such giant enhancement disappears for the $x$=0.4 sample when the FRM ground state is completely stabilized, i.e., no competing AFM phase at low temperature for the $x$=0.4 sample. The maximum photoinduced magnetization is around 40$\%$ of the saturated magnetization, tunable through pump fluence, Fig. \ref{fig4}(b). These features of optomagnets created from antiferromagnets can be useful in the pursuit of miniaturized magnets that can be integrated into spintronic devices to operate an on/off exchange coupling.

The giant optomagnet effect of the $x$=0.125 sample is initiated by a spin-flip absorption, followed by a significant enhancement due to the strong phase competition rather than the effective pulsed magnetic field from ultrafast lasers (photomagnetic pulses $H_{\text{eff}}$) \cite{Kimel2005Nature} which would require a strong spin-orbit coupling to create a virtual transition before spin flips to emit a magnon \cite{White1982PRB}. This can be verified via the comparison of $\Delta\theta_{\text{max}}$ versus pump fluence, i.e., $H_{\text{eff}}$, measured at 10 K and 20 K, Fig. \ref{fig4}(b). The hysteresis curves display a drastic difference between the two temperatures \cite{Kurumaji2015PRX} while $\Delta\theta_{\text{max}}$-$H_{\text{eff}}$ remains the same, allowing us to rule out the enhancement of the optomagnet effect due to the mechanism of photomagnetic pulses excitation \cite{Kimel2005Nature}.

Although, at the present stage, it is not crystal clear about how the initial spin flip gradually leads to the giant enhancement due to the complexity of many-body effects on the excited states, we can use the dynamic information gained from the $x$=0 sample to provide the following scenario. We consider the case of homogeneous Zn substitution on the T-sites. For the $x$=0.125 sample without magnetic field cooling at low temperature, the AFM phase is energetically preserved \cite{Kurumaji2015PRX,Nakayama2011JSPS}. Therefore, on average, one pair of Fe ions on the T-site with opposite moments should be replaced by nonmagnetic Zn ions per four magnetic bases, remaining zero magnetization \cite{ZnNote}. We can then consider the Zn doping as the removal of the pinning centers in the AFM state, that is to say, without doping the Fe sublattices are pinned to each other through the exchange interaction $J_{ij}$, with $J_{ij}$$<$0 being AFM coupling and $J_{ij}$$>$0 being FM coupling.

Next, we investigate the intralayer and interlayer magnetic couplings. Our ab~initio calculation determines the signs of the exchange interactions in the AFM phase before photoexcitation, as displayed in Fig. \ref{fig4}(c): (I) the intralayer nearest spin moments have $J_{\perp c}<$0, (II) the interlayer nearest O-sites and nearest T-sites have $J_{\text{oo}}<$0 and $J_{\text{tt}}<$0,  and (III) the interlayer nearest couplings have $J_{\parallel c}>$0  \cite{suppl}.

Knowing the detail of the exchange interactions without photoexcitation, we now focus on a potential scenario discussed below for the giant yet slow magnetization enhancement. Without Zn doping, we know that the optomagnet effect is small and its lifetime is relatively short, Fig. \ref{fig2}(b)-\ref{fig2}(c), for any pump fluence at any temperature. This can be understood as a result from strong exchange couplings of the AFM ground state and from the rather large difference in free energy between the AFM and FRM phase ($>$15 T at 20 K \cite{Kurumaji2015PRX}), both of which tend to relax the photoinduced magnetization. In the doped sample, Zn substitution suppresses such a pinning effect resulting from strong exchange interaction of the AFM ground state and reduces the free energy difference between the competing phases ($<$1.5 T at 20 K \cite{Kurumaji2015PRX}). When \Odd~occurs under a circularly-polarized pump, e.g., $\sigma^{-}$ that excites sublattice 1 in Fig. \ref{fig4}(d), the flipped spin moment (in blue color) creates new exchange interaction $J'_{\parallel c}$ of AFM coupling and $J'_{\text{oo}}$ of FM coupling, the dashed green double arrows in Fig. \ref{fig4}(d). Not only the new couplings favor the FRM state, but also the partial replacement of Fe with Zn decreases the pinning between its nearest neighbors, indicated by the dotted grey double arrows, stabilizing the new interaction without relaxing the excited spin moment in a relatively short time as that occurs in the sample of $x$=0. The third basis in Fig. \ref{fig4}(d) displays the photoinduced magnetic configuration favorable for the FRM phase. The subsequent slow enhancement, on a timescale of 100-200 ps, can be understood as further reorientation that occurs on the pumped sublattice, likely by slow spin-lattice relaxation \cite{SLNote}. After spin-lattice relaxation, spin temperature slowly elevates and the free energy difference further reduces to stabilize the photoinduced magnetization, while its polarity is already set by the initial circularly-polarized pump.

Finally, we find that spin-flip excitation from \Tdd~(see Fig. S7 in the supplemental material) displays exactly the same dynamics as that observed by using \Odd, including the similar timescale and maximum optomagnet effect, except the reversal of Kerr rotation sign under the same circularly-polarized pump, being consistent with the mechanism displayed in Fig. \ref{fig3}(c). From Fig. \ref{fig4}(e), we can see that \Tdd~ also creates $J'_{\parallel c}$ of AFM coupling and $J'_{\text{tt}}$ of FM coupling, as shown by the dashed green double arrows. However, to have the excited sublattices beneficial from the Zn substitutions, it demands the significant reduction in the strength of $J_{\text{tt}}$ after doping, so that pinning by the strong $J_{\text{tt}}$ exchange interaction can be reduced, as shown by the dotted grey double arrows. Therefore, the appreciable $J_{\text{tt}}$ coupling that occurs without doping will enable its reduction upon Zn substitution, allowing the subsequent spin flip from the pumped T-site that enormously enhances the optomagnet effect \cite{SLNote}. The giant optomagnet effect of the $x$=0.125 sample points to the importance of both intralayer coupling $J_{\perp c}$ (nearest neighbor coupling), and interlayer coupling $J_{\text{tt}}$ (next nearest neighbor coupling), which are responsible for the enhanced phase competition induced by Zn substitution on T-sites as supported by our ab initio calculation \cite{suppl}.

The critical finding in our study is the optomagnet effect in antiferromagnets \FMOX, with polarity switchable via a circularly-polarized pump without the need of a residual magnetization or a magnetic field. Our proposed mechanism indicates its generalization, applicable for many antiferromagnets in transition metal compounds, so long as the selective on-site $d$-$d$ transition is allowed, e.g., by having noncentrosymmetric structures. Furthermore, the major breakthrough we demonstrate is the enormous enhancement of the optomagnet effect via enhanced phase competition introduced by nonmagnetic Zn doping that energetically favors the photoinduced magnetization. Therefore, the operation is robust, the resulting magneto-optical Kerr effect is large, and the induced magnetization in such AFM materials has long lifetime of tens of nanoseconds up to a quarter millisecond. These novel properties of antiferromagnets, only accessible via light-matter interaction, may innovate the design and operation of spintronics as well as memory devices.

\section{Acknowledgement}
Y.M.S. thanks Prof. W. H. Chang for providing us the cryostat, temperature controller, and vacuum pump. Y.M.S. acknowledges Dr. C.-P. Chuu for the technical support of structure illustration. This research is granted by Taiwan Ministry of Science and Technology (MOST 104-2112-M-009-023-MY3, MOST 104-2738-M-009-006, and MOST 107-2628-M-009-004-MY3), and the Center for Emergent Functional Matter Science of National Chiao Tung University from The Featured Areas Research Center Program within the framework of the Higher Education Sprout Project by the Ministry of Education (MOE) in Taiwan. K.R.B. and G.Y.G. acknowledge the support from MOST 104-2112-M-002-002-MY3 and MOST 107-2112-M-002-012-MY3 as well as Academia Sinica Thematic Research Program (AS-TP-106-M07). T.K. and Y.T acknowledge the support from the JSPS Grant-In-Aid for Scientific Research(S) No. 24224009.

\section{Appendix A: Sample preparations and optical methods}
Single crystals of \FMOX~ were grown by chemical vapor transport, and the detail growth method was discussed in Refs. \onlinecite{Strobel1982JSSC,STROBEL1983JCG,Kurumaji2015PRX}.  We used (001) oriented samples to perform the optical experiment, which was based on an amplified laser system with a 5 kHz repetition rate and $\sim$200 fs pulse duration centered at 1030 nm. Photon energies of 0.5-2 eV were generated by optical parametric amplifiers. The spot size was $\sim$100-150 $\mu$m for various pump photon energies, and $\sim$40-50 $\mu$m for the probe. The pump fluence was kept near 4 mJ/cm$^{2}$ in the absorption measurements at room temperature. The probe fluence was kept $\sim$0.25 mJ/cm$^{2}$ at the photon energy $\sim$0.86 eV, to which the samples were nearly transparent so that the probe can measure pure MOKE signal without influence from absorption bleach. Except for the $x$=0.125 sample, all the data shown here were measured with pump fluence around 4 mJ/cm$^{2}$. All the measurements performed at low temperatures were cooled without magnetic field. Polarization sensitive detection was employed by using a balance detection to subtract S- and P-polarized signals, eliminating common signals that involve no polarization rotation. We calibrated the photoinduced magnetization by measuring the static Kerr rotation induced from the saturated magnetization under a magnetic field.  All the TRMOKE displayed here were obtained with a polar geometry, where the probe light is propagating along the optical axis without detecting any birefringence for $P6_{3}mc$ symmetry shown in Fig. \ref{fig1}(a).

\begin{thebibliography}{42}
\bibitem [{ZnN()}]{ZnNote}%
  \BibitemOpen
  \href@noop {} {}\bibinfo {note} {While this pair of Zn doping can either be
  both within one basis or separated averagely, the latter of which should turn
  out to be more favorable in our scenario as it has more Fe$^{2+}$
  unpinned.}\BibitemShut {Stop}%
\bibitem [{SLN()}]{SLNote}%
  \BibitemOpen
  \href@noop {} {}\bibinfo {note} {While further flip from non-excited
  sublattices through spin-lattice relaxation or through the exchange
  interaction on the excited state is also possible by following the
  reorientation of the excited sublattices, at the current stage we exclude the
  extent of reorientation that leads to the ground state of Ferri phase, as we
  do not observe any indication of domain nucleations in our measurement.
  Within the fluence range we explore, all the photoinduced signal can recover
  to the original state with repeatable size and timescale in Kerr rotation
  within the laser pulse repetition period ($\sim$200 $\mu$s).}\BibitemShut
  {Stop}%
\end{thebibliography}

%

\end{document}